\documentclass[12pt,floatfix]{JHEP3}
\usepackage{graphicx}
\usepackage{dcolumn}
\usepackage{bm}
\usepackage{amsfonts}
\usepackage{amsthm}
\usepackage{amsmath}
\usepackage{amssymb}
\usepackage{cite}
\usepackage{epsfig}

\newcommand{\bea}{\begin{eqnarray}}
\newcommand{\eea}{\end{eqnarray}}
\newcommand{\be}{\begin{equation}}
\newcommand{\ee}{\end{equation}}

\newcommand{\beq}{\begin{equation}}
\newcommand{\eeq}{\end{equation}}

\title{Neutralino Dark Matter from Indirect Detection Revisited}

\author{ 
 Phill Grajek\footnote{\tt{phillip.grajek@umich.edu}},  
 Gordon Kane\footnote{\tt{gkane@umich.edu}}, 
 Daniel J. Phalen\footnote{\tt{phalendj@umich.com}},
 Aaron Pierce\footnote{\tt{atpierce@umich.edu}},
 and Scott Watson\footnote{\tt{watsongs@umich.edu}}
\\ Michigan Center for Theoretical Physics\\
University of Michigan, \\ Ann Arbor, Michigan 48109, USA}

\abstract{We revisit indirect detection possibilities for neutralino dark matter, emphasizing the complementary roles of different approaches.  
While thermally produced dark
matter often requires large astrophysical "boost factors" to observe
antimatter signals, the physically motivated alternative of non-thermal
dark matter can naturally provide interesting signals, for example from
light wino or higgsino dark matter.  After a brief review of cosmic ray
propagation, we discuss signals for positrons, antiprotons,
synchrotron radiation and gamma rays from wino annihilation in the galactic
halo, and examine their phenomenology.  For pure wino dark matter
relevant to the LHC, PAMELA and GLAST should report signals.}

\keywords{Supersymmetry Phenomenology, Cosmology of Theories beyond the SM}

\preprint{MCTP-08-54}
\begin{document}

\section{Introduction}
While a host of astrophysical measurements have precisely determined the amount of dark matter in our universe, we do not yet know its identity.  At present one could imagine that the dark matter is a weakly interacting massive particle (WIMP), an axion, or something more exotic.  This situation should change, perhaps soon.  If the dark matter is indeed a WIMP, evidence for it could be found both at the Large Hadron Collider (LHC) and a host of dark matter detection experiments, both direct and indirect.

In this paper, we will assume that the dark matter is a WIMP, in particular the lightest supersymmetric particle (LSP).  
The identity of the LSP depends on the details of supersymmetry breaking.  Determining its identity will be a necessary step towards understanding the cosmological history of our universe, and an important clue towards the determination of the underlying theory. A phenomenologically attractive candidate is the lightest neutralino.  We concentrate on a case that is both physically well-motivated and potentially gives large signals for dark matter indirect detection: a non-thermally produced LSP with large annihilation cross section. This scenario does not require additional anomalously large astrophysical ``boost factors'' to produce interesting signals.

By now, a large literature on the indirect detection of dark matter exists.  For reviews, see \cite{JKG, HooperBaltz}.  We will place particular emphasis on a dark matter interpretation of positrons, for earlier work on this subject see, e.g., \cite{BaltzPositron,Kane1and2,Freese1and2,ProfumoUllio}.

For the LHC to provide complementary data on the dark matter\cite{Brhlik,PeskinBaltz}, it must be kinematically accessible.  Often, the dark matter is most efficiently searched for in the cascade decays of colored particles.  However, there can be a large gap between the dark matter mass and the lightest colored particle.  In models with gaugino mass unification, there is roughly a factor of seven between the WIMP candidate mass and the gluino mass.  In anomaly mediated models of supersymmetry breaking, the ratio is a factor of nine; in other models with non-universal gaugino masses, it can be a factor of a few.  Thus, if the gluino is to be produced copiously (say with a mass less than 2 TeV), the dark matter should not be too heavy.  In this paper, we will focus on a light mass region where the LSP is a wino with a mass of a few hundred GeV.

\subsection{Thermal vs. Non-Thermal Production}
Often, SUSY dark matter candidates are assumed to be produced from the primordial thermal plasma in the early stages of the universe (see e.g. \cite{JKG} for a review).
Under this assumption, the relic density of the LSP depends inversely on the annihilation cross section.  For a neutralino, $\chi$, one finds \cite{RPP,KolbTurner}:
\be \label{omega}
\Omega_{\chi} h^{2} \approx 0.1 \left(\frac {3 \times 10^{-26} \rm{cm}^{3} \rm{s}^{-1}}{\langle \sigma_{A} v \rangle}\right)
\ee
For the case of a light neutralino LSP (a few hundred GeV or less), this typically restricts the neutralino to have a substantial bino component as pure wino and Higgsino states (co)-annihilate very effectively to weak gauge bosons.  But precisely because of the smaller annihilation cross section, the annihilation signals from bino-like dark matter can be disappointingly small unless one appeals to large ``boost factors.''  This issue is further exacerbated by the fact that bino annihilations are p-wave suppressed in the early universe, and are thus suppressed by powers of the final state masses today. If, as is often the case,  the final state is $b$-quarks, the annihilation rate in our galaxy can be very small.
  
Models with gaugino mass unification often do typically give rise to a bino LSP, with its associated small annihilation cross section.  One is then challenged to reduce the relic density to the observed value.  However, if one does not assume a simple unification of gaugino masses at the high scale, other possibilities arise, well-motivated by top-down models of supersymmetry breaking.  One attractive possibility is a wino LSP.  This naturally occurs in theories where anomaly mediation gives the dominant contribution to the gaugino masses\cite{AnomalyMediation}.  It also occurs in string compactifications, see, e.g. \cite{Acharya:2008zi}.  This type of dark matter can also occur in the simplest models of split-supersymmetry \cite{Split,WellsPeV}, where the gauginos get anomaly mediated masses (with attendant loop suppression), but scalars receive large masses, suppressed only by the Planck scale.   

A light wino has a large annihilation cross section, which is good for indirect detection, but also implies a small thermal relic abundance.  The solution to recovering the correct relic abundance comes from non-thermal production.  Often, the very same models that predict a wino LSP also provide mechanisms by which the LSP is produced non-thermally.  If particles decay into the wino below its freeze-out temperature, this can provide the correct relic abundance\cite{WatsonKane}.  Excellent candidates for the decaying particle include gravitinos and weakly coupled moduli \cite{Moroi:1999zb,Giudice:2000ex,Acharya2008:bk}.  Non-thermal production of dark matter leads to WIMPs with larger cross sections, since the standard thermal relic abundance calculation no longer applies.  Since the flux of anti-particles coming from dark matter annihilations depends linearly on the cross section, this implies that non-thermal production of dark matter may lead to larger fluxes that may be detectable in future indirect experiments.\footnote{While we will concentrate on wino dark matter, the results are a bit more general.  In the region of interest, the winos annihilate nearly exclusively to $W$ bosons.  So, basically what we are probing is a dark matter candidate that annihilates to $W$'s with a given cross section.}

In the remainder of the paper, we review elements that enter any discussion of the indirect detection of dark matter.  First, we briefly review basics of cosmic ray propagation, as well as the form of the source term arising from dark matter annihilation. We then discuss constraints from both anti-protons and synchrotron radiation.  We then discuss prospects for observations of non-thermally produced wino dark matter in positrons and gamma rays.  With both PAMELA (a Payload for Antimatter Matter Exploration and Light-nuclei Astrophysics) and GLAST (Gamma Ray Large Area Space Telescope) in orbit, these two signals are particularly timely.  Throughout, we attempt to point out where astrophysical assumptions enter.  Finally, we comment on implications for the LHC, and briefly discuss implications for direct detection and indirect searches for dark matter via neutrinos.

\section{Cosmic Rays}

\subsection{Production \label{production}}
Our emphasis will be on the identification of high energy cosmic rays from dark matter annihilation.  However, disentangling this component relies on an understanding of the other components of cosmic rays.  Cosmic rays can be observed directly, e.g. from supernova ejecta (primaries). Alternately, these cosmic rays can interact with the interstellar medium producing secondaries.  Both components contribute to the cosmic ray background, and typically have a flux that is a power-law as a function of their kinetic energy.  This is an anticipated property of cosmic rays of astrophysical origin.  

The annihilation products of a dark matter particle will be associated with a given energy scale (its mass), and thus can conceivably be distinguished from power law backgrounds.  These annihilations will act as a source term:
\begin{equation}
Q = \frac{1}{2} \left(\frac{\rho(r)}{m_{\chi}}\right)^2 \langle \sigma v \rangle  \frac{dN}{dp}(p),  \label{eq:source}
\end{equation}
where $\rho(r)$ is the dark matter profile, and $\frac{dN}{dp}(p)$ is the spectrum of stable particles resulting from the annihilation.  We simulated $\frac{dN}{dp}(p)$ using PYTHIA \cite{Sjostrand:2006za} and altered the dark matter source code in GALPROP\cite{Strong:1998pw} to accept this as input.  

When looking at most indirect signals of dark matter, the profile of the dark matter is an important ingredient.  N-body simulations seem to favor cusped profiles at the center of the galaxy such as the Navarro-Frenk-White (NFW) \cite{Navarro:1996gj} and Merritt \cite{Merritt:2005xc} profiles, while dynamical observations of galaxies seem to favor cored profiles of the isothermal variety \cite{Gentile:2004tb}.  Current dark matter simulations do not yet include the effects of baryons.  Baryons dominate the gravitational potential in the center of our galaxy, so we find it prudent to consider three dark matter profiles. The first is the Navarro-Frenk-White profile:
\begin{equation}
\rho(r) = \rho_{\odot} \left(\frac{r_{\odot}}{r} \right) \left(\frac{1+\left(r_{\odot}/r_s \right)}{1+\left(r/r_s\right)} \right)^2,
\end{equation} 
with $r_s = 20$ kpc, where $r_{\odot} = 8.5$ kpc is the galactocentric distance of the sun and $\rho_{\odot} = 0.3$ GeV/cm$^3$ is the local dark matter density.  The second is the isothermal profile
\begin{equation}
\rho(r) = \rho_{\odot} \frac{1+\left(r_{\odot}/r_s \right)^2}{1+\left(r/r_s\right)^2} ,
\end{equation}
with $r_s = 3.5$ kpc, and finally the Merritt et al. profile 
\begin{equation}
\rho(r) = \rho_\odot \exp \left[-\left(\frac{2}{\alpha} \right) \frac{r^{\alpha} - r_{\odot}^{\alpha}}{r_s^{\alpha}} \right],
\end{equation}  
with $\alpha = 0.17$ and $r_s = 25$ kpc.

\subsection{Cosmic Ray Propagation}
Charged particles from dark matter annihilation must traverse part of the galaxy before arriving at detectors near Earth.  This propagation has a non-trivial effect on the form of the signal.

Annihilations will take place in both the galactic plane and the dark matter halo.  Once these particles are produced, they will either become confined by the galactic magnetic field to an approximately cylindrical region or escape the galaxy forever.  Their propagation may be described by a diffusion equation, whose details we will now review.  Some of the parameters entering this equation are uncertain, and will give rise to uncertainties in the observed dark matter signals.

\begin{figure}
  \begin{center}
    \scalebox{0.6}{\includegraphics{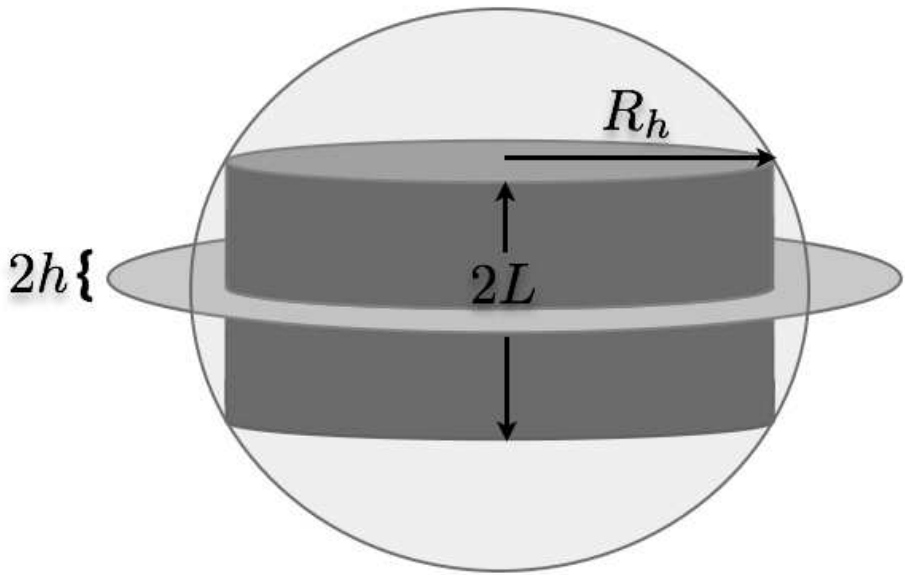}}
  \end{center}
  \caption{The diffusion zone (cylinder) is taken to have a height $2L$, with $L$ in the range of 4-12 kpc \cite{Strong:1999su}, whereas the radial direction is taken as $R_h = 20$ kpc (see figure 1). Most of the interstellar gas is confined to the galactic plane at $z=0$, which represents a slice through the cylinder and has a height of $2h=100$ pc.  Our solar system is then located in this plane at a distance of around $r_0=8.5$ kpc from the galactic center. All of this is enveloped by a spherically symmetric dark matter halo.\label{fig:cylinder}}
\end{figure}

In modeling propagation of cosmic rays through the galaxy, we will assume cylindrical symmetry (Fig. \ref{fig:cylinder}).  We will adopt a cylinder with height $2L$, and some maximum radius $R$.  The stars and dust will be confined to the galactic plane $z=0$.  The dark matter halo has a spherical symmetry.  The particles are allowed to freely escape at the boundaries, and propagation within the cylinder is described by the diffusion-loss equation \cite{Strong:1998pw}: 
\begin{eqnarray}
\frac{\partial}{\partial t}\frac{dn}{dp}(\vec{x},t,p) &=& \vec{\nabla}\cdot (D_{xx}(\vec{x},E,t)\vec{\nabla}\frac{dn}{dp}-\vec{V}\frac{dn}{dp})-\frac{\partial}{\partial p}(\dot{p}\frac{dn}{dp} - \frac{p}{3}(\vec{\nabla}\cdot \vec{V})\frac{dn}{dp}) \nonumber \\
&& + \frac{\partial}{\partial p}(p^2 D_{pp} \frac{\partial}{\partial p} (\frac{1}{p^2}\frac{dn}{dp}))+Q(\vec{x},t,p). \label{eq:diffeqn}
\end{eqnarray}

{\it The Diffusion coefficient:} Cosmic rays diffuse out of the galaxy by scattering off inhomogeneities in the magnetic field.  The diffusion coefficient
\begin{equation}
D_{xx} = \beta K_0 \left( \frac{{\mathcal R}}{{\mathcal R}_0} \right)^{\delta},
\end{equation}
is a function of the rigidity ${\mathcal R} \equiv p/Z$ where $Z$ is the atomic number.  $K_0$ is a constant, ${\mathcal R}_0$ is some reference rigidity, $\beta$ is velocity, and $\delta$ is the scaling with respect to the momentum.  We take a default value $K_0=5.8 \times 10^{28}$ cm$^2$ s$^{-1}$. The scaling, $\delta$, is set by the spectrum of magneto-hydrodynamic turbulence in the interstellar medium.  It is $0.33$ for a Kolmogorov type spectrum, and $0.5$ for a Kraichnan type spectrum\cite{Ptuskin:2005ax}.  Values in this region are reasonable.  The dependence on $\beta$ can be understood simply: higher $\beta$ increases collisions with the inhomogeneities, and hence the diffusion.   

{\it Energy Loss:} The energy loss, $\dot{p}$, comes from several sources: bremsstrahlung, Coulombic interactions with ionized gasses, inverse Compton scattering with starlight and with the CMB, and synchrotron radiation.  Inverse Compton scattering and synchrotron radiation are the largest contributors to energy loss for electrons and positrons and not important for anti-protons.  In the case of electrons the energy loss time is sometimes parametrized by $\tau$, with $\dot{p} \propto p^2/\tau$.  A typical value is $\tau \approx 10^{16}$ sec.

{\it Re-acceleration:} Re-acceleration comes from second order Fermi processes and is described as diffusion in momentum space.  It enters the diffusion equation via the term proportional to $D_{pp}$ in Eqn.~(\ref{eq:diffeqn}).  If magnetic fields move randomly in a galaxy, cosmic rays can be speed up when reflected from a magnetic mirror coming them.  Likewise, they are slowed down by reflecting from a mirror moving away. The diffusion coefficient $D_{xx}$ and the re-acceleration coefficient $D_{pp}$ are related via the Alfv\'en velocity \cite{SeoPtuskin}.  These magnetic field waves are moving slowly with respect to higher energy cosmic rays, so re-acceleration only will effect the low energy cosmic rays.

{\it Convection:}  The convection current $\vec{V}$ can be thought of as a wind streaming in the $z$ direction outward from the galactic plane.  It is due to the outgoing plasmas from the galaxy, and in our galaxy can be thought of as coming from cosmic rays accelerating the plasma \cite{Zirakashvili}. For the case of positrons, convection and annihilations in the disk can be neglected.

{\it Source terms and radioactive decays:} For astrophysical sources, the source term $Q$ is expected to proportional to a power law $\propto p^{-\gamma}$ localized in the galactic plane.  It may also contain sources and sinks due to unstable cosmic rays.

We will employ GALPROP \cite{Strong:1998pw} for numerical solutions to the diffusion-loss equation.

\subsection{Some Uncertainties}
Measurements of the boron to carbon ratio help to fix the ratio of primary to secondary cosmic rays. Boron is produced purely as a secondary, while carbon is mostly primary.   This observation helps fix both the height of the diffusion zone and the diffusion parameters $K_0$ and $\delta$. However, there can exist a large degeneracy between these parameters\cite{Donato:2003uz,Bergstrom:2006tk}. Increasing the height of the diffusion zone traps more cosmic rays.  This can be compensated by a simultaneous change in the diffusion parameter that allows cosmic rays to quickly escape the galactic plane.  Since anti-protons of a non-dark matter origin are produced in the galactic plane as secondaries, just as boron is, this apparent degeneracy of parameters does not give rise to a large uncertainty in the background prediction.  Once the primary flux of protons is fixed (measured), the B/C ratio gives a rather precise prediction for the (astrophysical) anti-proton flux.

Unfortunately, the dark matter does not share the same independence of the astrophysical parameters.  Depending upon which set of diffusion parameters are chosen, different dark matter signals result.  The reason is that dark matter annihilations are not confined to the galactic plane.  Rather, they occur throughout the halo, and  increasing the diffusion zone includes more primary cosmic rays from dark matter.  This change in $L$ is not completely compensated by an increase in the diffusion out of the galactic plane as in the case of the background. Moreover, this increase in the height of the diffusion zone will affect positrons and anti-protons differently, as we will discuss in the following sections.

\section{Experimental Constraints on Non-thermal Neutralinos}
In this section  we use GALPROP \cite{Strong:1998pw} to numerically solve the propagation equation (\ref{eq:diffeqn}) and find the expected flux of positrons and anti-protons, as well as the synchrotron radiation coming from the annihilation products of neutralino dark matter.  When appropriate, we have checked these results explicitly using DarkSUSY \cite{Gondolo:2005we}, and found similar results for similar values of the astrophysical parameters.  We discuss the possibility of neutralino dark matter annihilations to explain an excess of positrons as suggested by the HEAT \cite{Barwick:1995gv,Barwick:1997ig} and AMS-01\cite{AMS01} data, while simultaneously respecting the observed flux of anti-protons as measured by BESS \cite{Orito:1999re}.  At present, the anti-protons do not show any peculiar spectral features (though their flux is perhaps somewhat lower than expected).  We use this data to set bounds.  We also discuss bounds on the neutralino annihilation cross section from synchrotron radiation in the ``WMAP haze'' \cite{Finkbeiner:2003im, Finkbeiner:2004us, Hooper:2007kb} obtained from the WMAP3 data \cite{Spergel:2003cb}, and discuss implications for the GLAST experiment.

\subsection{Anti-Proton Bounds}

\begin{figure}[h]
  \begin{center}
    \scalebox{0.6}{\includegraphics{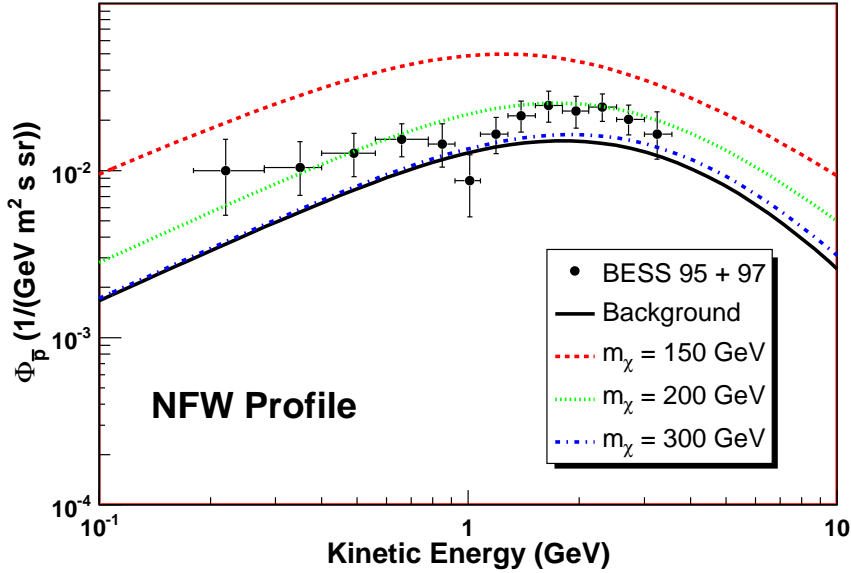}}
  \end{center}
  \caption{The flux in anti-protons for varying neutralino mass ($m_{\chi}=150,200,300$ GeV).  We have taken a diffusion zone height of $L=4$ kpc. \label{protonfluxmass}}
\end{figure}

Before attempting to fit the HEAT data (or make predictions for the PAMELA experiment), we must take into account bounds from anti-protons. We will compare to the BESS 95 + 97 data \cite{Orito:1999re} taken at the solar minimum, and modulate the interstellar spectrum with a potential of 550 MV.  More recent data from both the 1998 BESS data\cite{BESS} and the BESS-Polar data \cite{Abe:2008sh} will have a different modulation potential but display the same trends.  In Figure \ref{protonfluxmass}, we show the anti-proton flux for varying mass of the wino-like neutralino.  As expected, increasing the mass of the wino pushes the spectrum to slightly harder energies.  The dominant effect, however, is that an increase in the wino mass results in a decrease in the annihilation cross section as well as number density in the profile, which changes the overall normalization of the curve. Apparently, a wino mass of 150 GeV gives too high a flux, but 200 GV is (marginally) consistent. 
   
\begin{figure}[h]
  \begin{center}
    \scalebox{0.6}{\includegraphics{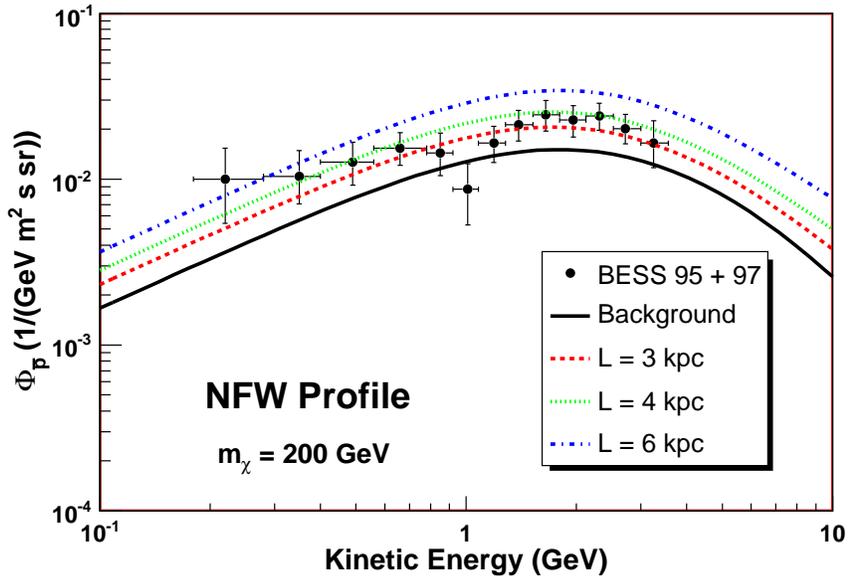}}
  \end{center}
  \caption{The flux in anti-protons for varying height of the diffusion zone cylinder with and NFW dark matter profile. We have taken a $m_{\chi}=200$ GeV wino.\label{protonfluxdiffusion}}
\end{figure}

As can be seen in Figure \ref{protonfluxdiffusion}, these constraints are sensitive to the diffusion zone height.  Here, we fix the neutralino mass at  200 GeV, and vary the diffusion height, $L$.  Clearly the diffusion height directly affects the anti-proton flux.  Again, we see that for a height of $L=4$ kpc, $m_{\chi}=200$ GeV is accommodated by the anti-proton data, but for larger diffusion cylinders, heavier winos would be required to be consistent with the anti-proton data.

\begin{figure}[h]
  \begin{center}
    \scalebox{0.6}{\includegraphics{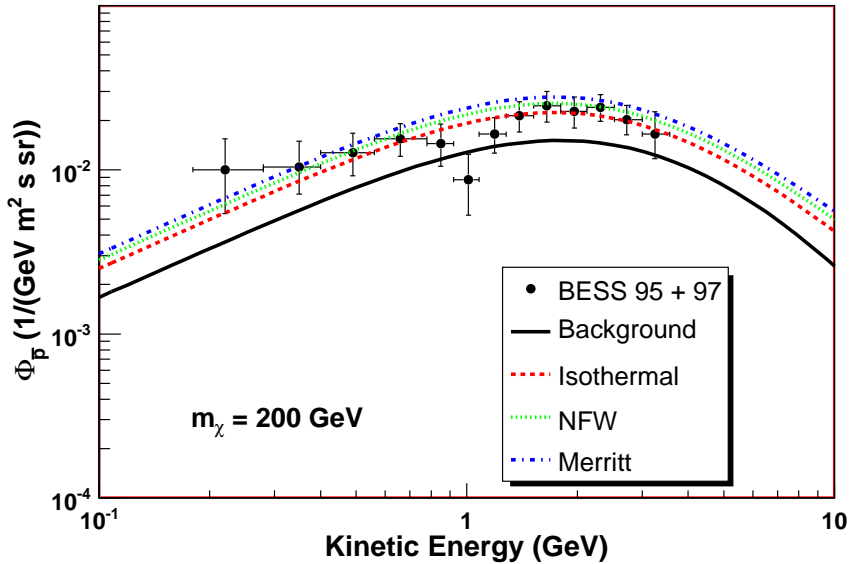}}
  \end{center}
  \caption{The flux of anti-protons is shown using different dark matter distributions.  We have fixed $L=4$ kpc, and the wino mass to be $m_{X} =200$ GeV. Since the anti-protons may sample the inner region of the galaxy, the cuspiness of the profile does effect the anti-proton flux.}
  \label{fig:pbarprofile}
\end{figure}

This minimum allowed wino mass is also a function of the dark matter distribution in the galaxy.  Because anti-protons do not lose energy very efficiently (relative, to say, positrons), they come to us from a large region, and can potentially sample the inner portion of the galaxy, where the dark matter distribution can vary dramatically among different choices of profile.  To assess the dependence of the profile on potential dark matter flux from anti-protons, we varied the profile in Fig.~\ref{fig:pbarprofile}.  Note that going from an NFW profile to another profile changes the flux of anti-protons from the dark matter particle by roughly $\pm 15\%$.

Our investigation of the anti-proton flux indicates that a pure wino of approximately 200 GeV is consistent with the data.  To achieve significantly lower masses, one would have to push the astrophysical uncertainties.  A 150 GeV pure Higgsino, however, is consistent with the data.  At this mass, its annihilation cross section is approximately one order of magnitude below that of the wino. 

\subsection{Synchrotron Radiation}
An excess of synchrotron radiation in the WMAP three year data \cite{Spergel:2003cb}, particularly significant for angles south of the galactic plane, has been suggested by subtracting out known foregrounds\cite{Finkbeiner:2003im, Finkbeiner:2004us, Hooper:2007kb}. The residual component has a harder spectrum than other known sources for microwave emission, and has been dubbed the WMAP haze.   Thus it seems that there is an unknown source of relativistic electrons and positrons moving in the galactic magnetic field, contributing to synchrotron emission. These electrons and positron could potentially come from dark matter \cite{Finkbeiner:2004us}.  

While the exact interpretation of the haze is unclear at present, at minimum one should at least check that any potential dark matter candidate does not super-saturate the amount of synchrotron radiation.  This has been noted by Hooper \cite{Hooper:2008zg}, who uses this observation to potentially place bounds on dark matter candidates.  Here, we briefly revisit these bounds and semi-quantitatively discuss the astrophysical uncertainties that enter them.\footnote{It should be noted that very strong bounds from $X$-rays might result if strong $B$-fields exist near the black hole near the galactic center\cite{UllioMulti}.  We do not pursue these bounds further here.}

First, we discuss the particles that contribute to the WMAP bands.  These electrons have energy greater than 5 GeV.  This can be shown by analyzing the equation for synchrotron emission.  We use the formula of \cite{Ghisellini:1988}, 
\beq
\epsilon_S(\nu, \gamma) = \frac{4\pi\sqrt{3}e^2 \nu_B}{c} x^2 ( K_{4/3}(x) K_{1/3}(x) - \frac{3}{5} x (K_{4/3}^2(x) - K_{1/3}^2(x))) \label{eq:synch}
\eeq
with
\beq
x = \frac{\nu}{3\gamma^2 \nu_B},
\eeq
$\gamma$ is the boost factor, and the critical frequency is $\nu_B = e B/2\pi m_e c$.  Here, $K_n$ is the modified Bessel function of order $n$.  This formula gives the synchrotron emission of the electron into all angles, averaged over an isotropic pitch angle distribution of the electrons with the magnetic field.  

\begin{figure}
  \begin{center}
    \scalebox{0.9}{\includegraphics{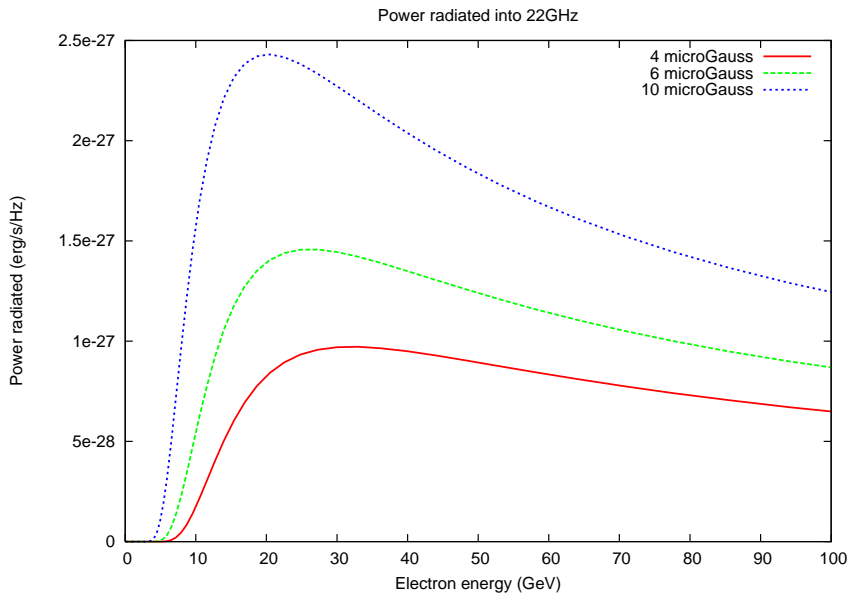}}
  \end{center}
  \caption{Power radiated into 22 GHz as a function of electron energy for different values of the galactic magnetic field.  Notice that for energies below 5 GeV, there is negligible radiation. \label{fig:powerradiated}}
\end{figure}

Figure \ref{fig:powerradiated} shows the amount of synchrotron radiation into the 22 GHz band as a function of the electron energy for a few different values of the magnetic field.  This band is observed by WMAP, and it gives the most statistically significant contribution to the haze.  Error bars in other bands are larger.  Emission from energies below 5 GeV is negligible.  This demonstrates the link between the haze and high-energy electrons and positrons.  Thus, the excess in the HEAT data and the synchrotron emission can be linked to the relativistic electrons of similar energy.  Indeed, any positron excess from a future experiment will potentially contribute to the haze at some level. If both the haze and positron excess arise from dark matter, then reconciling them will probe the astrophysical parameters of our galaxy.

As a point of reference, \cite{Hooper:2008zg} argues that a pure wino that gives the full dark matter abundance would be excluded by the haze unless its mass exceeds 700 GeV. This is a very strong bound, and as we will see, would largely preclude any interpretation of any current or future excess in positrons as simple supersymmetric dark matter.

Central to placing this bound is an understanding of how electrons and positrons lose energy within the galaxy.
This is controlled by the relative importance of the radiation field and magnetic field in the region of interest.  Large magnetic fields will cause the energy loss due to synchrotron radiation to dominate (and hence yield strong bounds from the haze).  Large radiation fields will cause inverse Compton scattering to dominate.  Reference \cite{Hooper:2008zg} assumes a relationship between the energy density in the magnetic field and in the radiation field as: $U_B / (U_B + U_{rad}) \sim 0.26 $.  With a naive equipartition relation one would find this ratio $\sim 0.5$.  There is no tight argument for equipartition between these two contributions.  However, it is not unlikely that this relation should roughly hold at least approximately.  After all, the $B$-field is related to cosmic rays, whose source is astrophysical objects.  These, in turn,  should roughly trace that radiation distribution.\footnote{We thank Dan Hooper for discussion of this point.}  

Having argued that the bound will sensitively depend on the choice of the magnetic and radiation field, we set about to semi-quantitatively investigate this effect by using a different initial set of assumptions.  Our view is that our starting point is not obviously less motivated than that of \cite{Hooper:2008zg}. Our results might then give some indication of the size of the astrophysical uncertainties. Alternately, if one wishes to have a light dark matter particle with large cross section, our discussion will tell you what properties the galaxy must  have to accommodate such a candidate. 

To find the synchrotron sky map arising from our dark matter annihilation, we use GALPROP 50.1 \cite{Strong:1998pw} for the propagation of our dark matter derived electrons. We use the parameters $K_0 = 5.8 \times 10^{28}$  cm$^2$ /s, $\delta = 0.5$ (consistent with a Kraichnan spectrum of interstellar turbulence) \cite{Ptuskin:2005ax}, and L = 4 kpc, but find our results are relatively insensitive to these choices.  Other choices for propagation parameters yield changes of roughly 10\% in the results.   The energy loss term is set by the relativistic Klein-Nisha cross section of cosmic rays on the interstellar radiation field combined with the synchrotron radiation from the magnetic field.  The injection spectrum of dark matter is modified to accept input from PYTHIA 6.4 \cite{Sjostrand:2006za}.  Following \cite{Hooper:2007kb}, we average emission over 20 degrees in longitude.  For the interstellar radiation field, we use the fields from \cite{Porter:2005qx, Moskalenko:2005ng} that are provided with the GALPROP package.  We model the magnetic field by an exponential decay
\begin{equation}
B(r,z) = B_0 e^{-|r|/r_0 -|z|/z_0}. \label{eq:bfield}
\end{equation}
We chose the characteristic distance $r_0$ such that the local magnetic field is $3\, \mu G$, and chose $z_0$ such that the field falls off quickly away from the galactic plane that is supposed to be responsible for creating this field.  Also, we will use equation \ref{eq:synch} to find the synchrotron radiation.  With sky-map in hand, following the same approach as \cite{Hooper:2008zg}, we use the synchrotron data of \cite{Finkbeiner:2004us} to constrain possible dark matter candidates.  Again, we do not assume a thermal history, and instead impose that our dark matter candidates make up all the relic density by fiat.  We find a 90\% confidence level upper bound on the annihilation cross section by using a $\chi^{2}$ fit, allowing the addition of a constant background synchrotron piece, independent of angle from the galactic center (relating to possible uncertainty in the subtraction procedure of Finkbeiner, et al.).  \footnote{Unlike \cite{Hooper:2008zg}, we impose the fit over the entire interval from 5 to 35 degrees south of the galactic plane.}

It should be noted that we do not recalculate the residual haze for each choice of the magnetic field.  However, since the approach of \cite{Finkbeiner:2004us} was simply to derive the haze by doing a comparison of sky-maps close to and away from the core, we view this as a reasonable first approximation.

\begin{figure}
  \begin{center}
    \scalebox{1}{\includegraphics{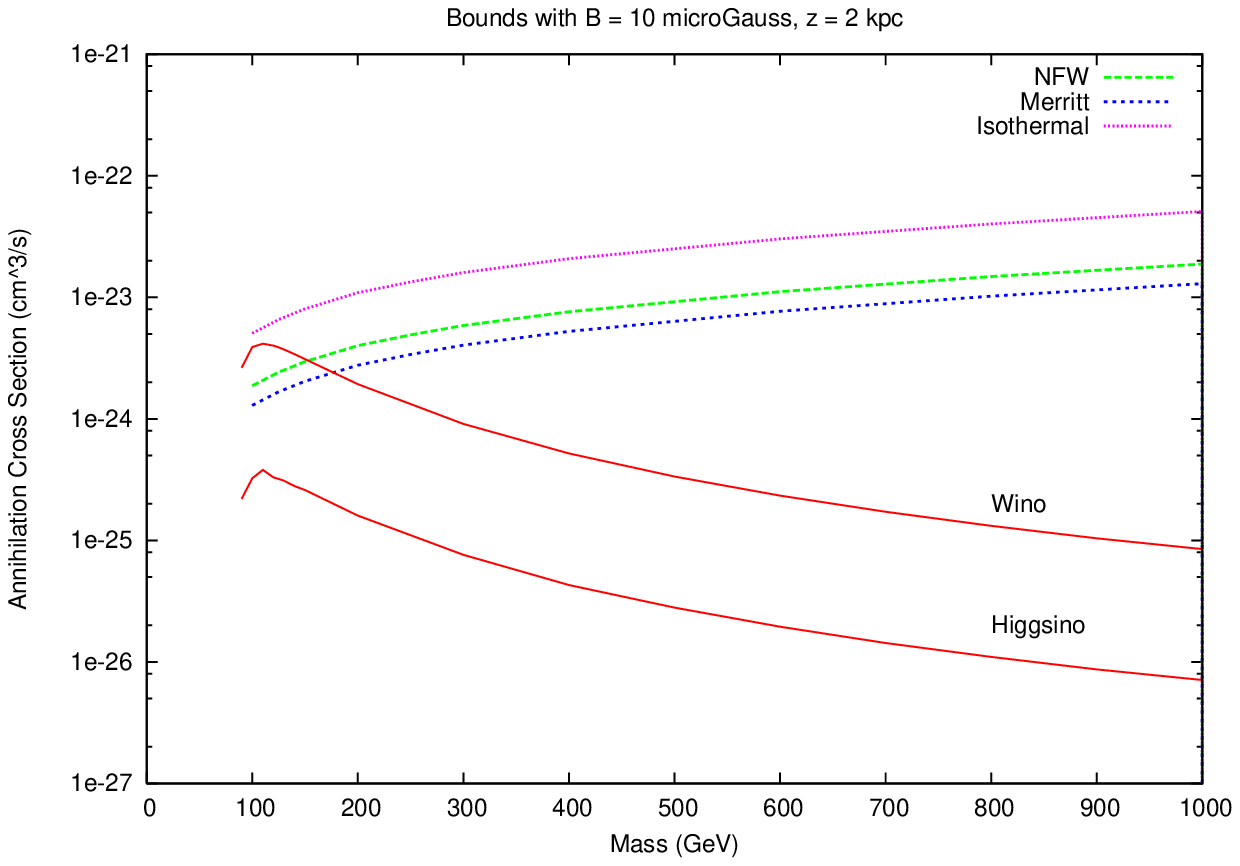}} 
    \scalebox{1}{\includegraphics{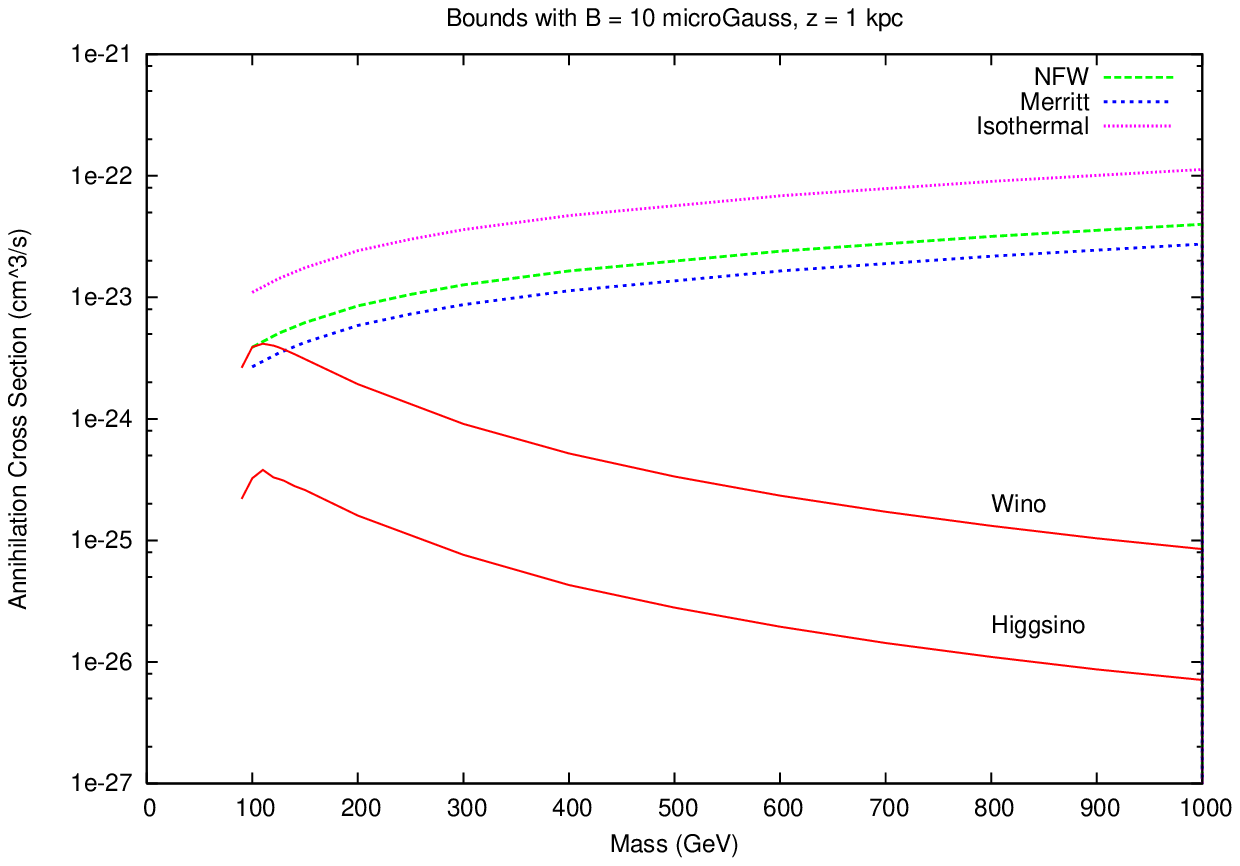}}
  \end{center}
  \caption{Bounds on the annihilation cross section into $W^+W^-$ from synchrotron radiation.  We have used the propagation parameters described in the text and only vary the magnetic field properties here.  
\label{fig:synchbounds}}
\end{figure}

For a cuspy profile, most of the dark matter annihilations will happen in the galactic core.  These then propagate outward until they are in the region we are looking at, 1 - 3 kpc from the center.  They then radiate into the frequency band observed.  Taking the approach outlined above, with $z_{0} = 2$ kpc, we find the results in the top panel of Fig.~\ref{fig:synchbounds}. In particular, for a pure wino, for an NFW profile we find the bound of 300 GeV, much less stringent than the original bounds from \cite{Hooper:2008zg}.  This is dominantly due to our choice of radiation field maps \cite{Porter:2005qx, Moskalenko:2005ng}.  For these maps, $U_B / (U_B + U_{rad}) \sim 0.1$ for $B_0 = 10$  $\mu G$ in the inner few kpc.  A larger value for this ratio pushes us towards the limits of \cite{Hooper:2008zg}.    If an even smaller $B$ field were present, near the galactic center, perhaps as small as 5 $\mu G$,   this would further degrade the limits to the point where the bounds from anti-protons become competitive with (or exceed) these bounds.

Finally, we briefly discuss the effect of the $z$ profile of the magnetic field.  It is not clear exactly what form the $z$ dependence of the $B$ field should take.  Taking $z_0$= 1 kpc again loosens the bound relative to our default choice of $z_{0}$= 2 kpc.  This is shown in the lower panel of Fig.~\ref{fig:synchbounds}.  Here, the bound on the pure wino dark matter only excludes 125 GeV wino dark matter, even for the relatively peaked NFW profile.  In short, the local $B$-field (i.e. where synchrotron radiation is being measured) has a large effect on the size of the synchrotron radiation signal.

Figure \ref{fig:synchbounds} also shows the dependence on the galactic profile.  Those that have a steeper rise towards the galactic center will give a larger contribution to synchrotron radiation.  If the profile is somewhat softer than NFW then the bound is further weakened (this effect was also very clearly shown in \cite{Hooper:2008zg} where a flat and NFW profile were shown).  If the less-peaked isothermal profile is chosen, for example, all bounds due to synchrotron radiation are eliminated, even in the case where the $B$ field falls off with $z$ relatively slowly.

Also shown in the figure are the annihilation cross sections for pure wino and pure Higgsino at low velocity.  For masses above $M_{W}$, both types of dark matter will annihilate almost exclusively to $W$ bosons.  Thus, discussions of $\gamma$-rays, synchrotron, $\bar{p}$ and positron signals will be identical for wino and Higgsino dark matter of the same mass, once this cross section difference is accounted for.    

There is a very clear relationship between the halo profile and what types of experiments are best suited to look for dark matter.  If the halo is quite peaked towards the center of our galaxy, then experiments that look for photons from this region, either gamma rays or synchrotron, will be best suited to find the dark matter.  If, however, the dark matter distribution rises more slowly, then it is no longer clear that the center of the galaxy is the best place to look. Indeed, one can then look for electrons and positrons directly (perhaps from annihilation to W bosons), rather than looking for indirect by-products of annihilation (synchrotron, or continuum gamma rays).  We now discuss this possibility.

\subsection{Positrons}

In the case of positrons, it is useful to consider the positron fraction, which includes both the primary flux of positrons $\Phi^{prim}_{e^+}$ as well as the background $\Phi^{sec}_{e^+}$ and the analogous fluxes for electrons, i.e.
\be
\Phi=\frac{ \Phi_{e^+}^{prim}+\Phi_{e^+}^{sec} } {\Phi^{prim}_{e^+}+\Phi^{sec}_{e^+}+\Phi^{prim}_{e^-}+\Phi^{sec}_{e^-}},
\ee
as this ratio allows for  cancellation of systematic errors and the effects of solar modulation (if we assume no charge bias).  Preliminary indications from PAMELA data \cite{PAMELA} indicate, however, that this charge bias may be important for low energies.  Since the dark matter signals we will consider will involve production of electron and positrons at multi-GeV energies, we believe charge bias should be safely negligible in this regime.

\begin{figure}
  \begin{center}
    \scalebox{0.6}{\includegraphics{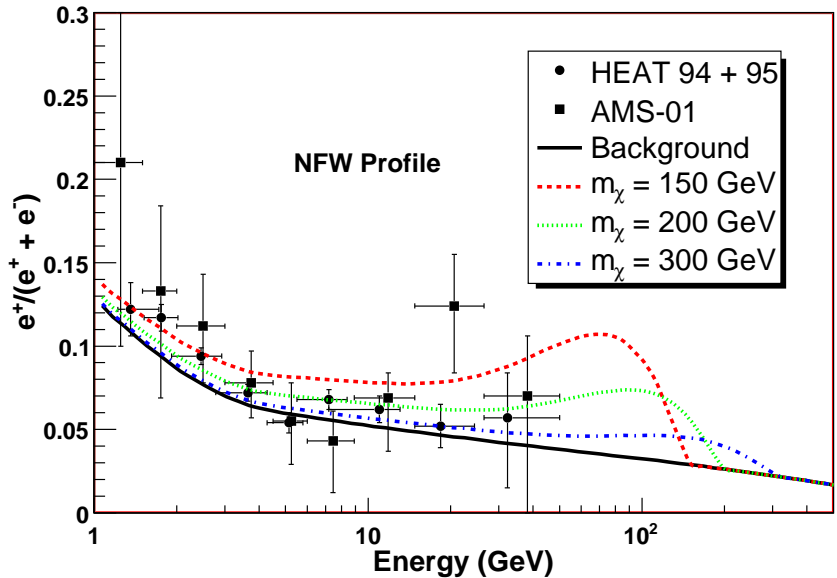}}
  \end{center}
  \caption{The positron flux from annihilations of non-thermally produced wino-like neutralinos for varying masses ($m_{\chi}=150,200,300$ GeV) keeping and NFW profile.  We have also included the data from the 1994-95 HEAT balloon based observations \cite{Barwick:1995gv,Barwick:1997ig}  and measurements from AMS-01 \cite{AMS01}. \label{posfluxvarymass}}
\end{figure}

In Figure \ref{posfluxvarymass}, we consider a purely wino-like neutralino for masses in the range $150-300$ GeV.  We have also included in the figure the data from the 1994 and 1995 HEAT missions \cite{Barwick:1995gv,Barwick:1997ig}, as well as the data from AMS-01 \cite{AMS01}.  The background curve is generated using the parameters of \cite{Cholis:2008vb}, with an Alfv\'en velocity of 20 km/s. At present the data begins to deviate from the background curve around 10 GeV, though the error bars are still large.  The error bars should shrink dramatically with new data from PAMELA, at which point one might attempt to fit the data with a WIMP signal.

One might be able to determine the mass of the WIMP from this data.  We see that the spectrum peaks slightly below $m_{\chi}/2$. This arises from annihilation to W-boson pairs and then subsequent decays to $e^+ / e^-$ near threshold.  At present, there is no turn-over in the data.  If PAMELA sees a turn-over in the data, then this would make a indirect measurement of the WIMP mass.  A pure wino of up to 400 GeV might be eventually observed by PAMELA (see \cite{ProfumoUllio,MasieroUllio}).

We find similar results for neutralinos that contain some bino or Higgsino component in addition to the wino, however in the case of the bino-like neutralino this can not be too large, otherwise the dark matter will not make a large contribution above the background.   

For the case of $\bar{p}$ and synchrotron radiation, there were important astrophysical uncertainties.  In particular, the distribution of of dark matter in the halo had a strong effect on the synchrotron bounds.  The size of the cylindrical region to which the dark matter annihilation products are confined by the galactic magnetic field has a large effect on the $\bar{p}$ flux.

These two systematics have a much smaller effect on the signal from positrons.  The reason is that positrons come from nearby: the typical diffusion length is only a few kpc.  Errors in the background are typically much larger than the differences induced in the signal by astrophysical uncertainties.  In this section we adopt the NFW halo profile as our canonical choice, noting that we find no significant changes for other profiles. Changing the height of the diffusion cylinder also does not have a very large effect on the positron ratio.  We investigated the same cylinder height as shown in the anti-proton section, and again found variations that were small when compared with other uncertainties in the astrophysical backgrounds. 

\begin{figure}
  \begin{center}
    \scalebox{0.6}{\includegraphics{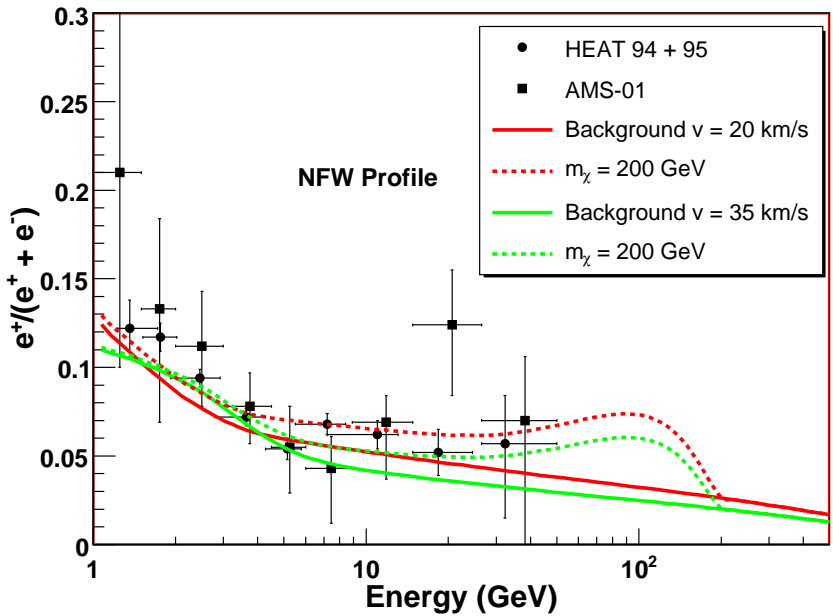}}
  \end{center}
  \caption{The positron ratio is shown for two different background curves, with (dashed) and without (solid) a Dark Matter contribution.  The two background curves correspond to different values of the Alfv\'en velocity, $v= 20$ km/s (red/dark) and $v= 35$ km/s (green/light).} \label{posfluxvarybkgrd}
\end{figure}

Re-acceleration can have an effect on the positron signal, however.  In Figure \ref{posfluxvarybkgrd} that using different backgrounds compatible with B/C will vary the positron signal as well.  We have used backgrounds with varying Alfv\'en velocities from \cite{Cholis:2008vb}.  The change in Alfv\'en velocity affects the low energy spectrum.  Once the low energy background is normalized to data, this affects the prediction at high energies.

In all our discussions, we have not assumed any ``boost factor''. If there is some additional substructure in the halo within the diffusion length of the positrons, it is possible to enhance this signal somewhat.  However, this substructure could also affect the signal in anti-protons, though not in a precisely identical way as they have a different diffusion length.

\subsection{Gamma Rays}
Dark matter annihilations can produce gamma rays in two ways: either via continuum production through decays of pions in hadronic decay products or directly.  Typically, direct annihilations to gamma rays are loop suppressed.   Here, we will concentrate on the continuum gamma ray flux, expected to be most relevant for GLAST. 

Since photon signals are independent of the propagation parameters for cosmic rays in the galaxy, the sky-map of gamma rays from dark matter annihilations directly trace the density profile.  The flux of gamma rays is given by 
\begin{equation}
\Phi_{DM}(E,\psi) = \frac{1}{2} \frac{\langle \sigma v \rangle}{4 \pi }  \int_{l.o.s.} dl(\psi) \frac{\rho(l)^2}{m^2_{\chi}} \sum_i \frac{dn_{\gamma i}}{dE} B_i,
\end{equation}
where the sum is over the different annihilation channels.  $B_{i}$ is the branching fraction;  $\frac{dn_{\gamma i}}{dE}$ is the gamma ray yield, and the integral is over the line of sight.  In our calculation of the continuum gamma rays from pions, we take $B_{i} =1$ for the $W^{+} W^{-}$ final state and zero, otherwise.

EGRET has looked at these signals in the 30 MeV - 50 GeV range, and has found no signals (see however \cite{deBoer}).  GLAST has recently launched and will have an increased sensitivity over this range.  It will also extend observations to 300 GeV.  Recently, \cite{Baltz:2008wd}, updated the work of \cite{HooperEGRET}, and studied of the sensitivity of GLAST to different dark matter models.  Here we discuss the implications of these studies for non-thermal wino dark matter.

The results of Fig. 6 of \cite{Baltz:2008wd} can be interpreted to place bounds on light winos.  In particular, in the case of an NFW profile, wino masses below $\approx$ 300 GeV should have already been observed by EGRET.  For an isothermal profile, the bounds are much weaker, less than 150 GeV.  In this case, the strongest bound at present comes from the anti-proton flux.

Given the relatively small masses found in the previous section necessary to explain the HEAT or AMS-01 data (or a future large excess at PAMELA), it is fair to say that there is already some tension between positron signals (if interpreted as dark matter)  and the absence of a signal in $\gamma$ rays if the profile is NFW (or cuspier).  If the profile is somewhat softer than NFW, however, then it is possible to accommodate both results.  A reduction in the astrophysical factor
\begin{equation}
J \equiv \int_{l.o.s.} dl(\psi) \frac{\rho(l)^2}{m^2_{\chi}},
\end{equation}
by a factor of $\approx$ six below its NFW value is necessary for EGRET to accommodate a 200 GeV wino and has no effect on positrons.  The improved sensitivity of GLAST suggests that an observation in gamma rays is in fact likely for such a WIMP.  The study \cite{Baltz:2008wd} suggest that a pure wino up to 500 GeV could be observed by GLAST at 3$\sigma$ after 5 years of running, assuming an NFW profile.  

Here, we have focused exclusively on the bounds (and potentially signals) coming from our own galaxy.  This approach will depend on the ability to successfully subtract away point sources and other diffuse backgrounds\cite{HooperDodelson}.  To avoid these sources, dwarf galaxies might be competitive places to look depending on their dark matter profiles.

\section{Comments and Conclusions }
A non-thermally produced wino is a well-motivated candidate for the dark matter observed in our universe.  Its large annihilation cross section could potentially allow it to explain the suggestion of an excess from HEAT/AMS-01, which could be confirmed soon by PAMELA.  However, to avoid conflict with bounds from gamma rays (and perhaps synchrotron radiation), the dark matter distribution cannot be too highly peaked towards the center.  There is already some tension in the case of an NFW profile.  This fact suggests that if PAMELA were to observe an excess in positrons that comes from dark matter, GLAST should follow with a confirmation.

Any candidate detection by PAMELA and GLAST will need to be examined in the context of direct detection experiments.  We do not do that here, since the pure wino LSP suggested by the present positron excess gives signals well below the current sensitivity of the current direct detection experiments. However, adding an admixture of Higgsino to the neutralino allows an increase in the direct detection cross section (via the $\tilde{w}-\tilde{h}-h$ coupling).  An increased Higgsino content also increases the capture cross section on the sun, allowing for a possible indirect detection via neutrinos. Thus, signals in these types of experiments could help to probe the Higgsino content of the LSP.

At the LHC, a pure wino of a few hundred GeV by itself may be difficult to observe via direct production.  However, it may be possible to find it in decays or associated production.  The sensitivity of this modes depends on the mass of the lightest colored mode.  In minimal models of anomaly mediation \cite{AnomalyMediation}, the ratio of the wino mass to the gluino mass is a factor of nine.  So a 200 GeV wino implies a 1.8 TeV gluino, which might preclude an early discovery.  However, if the mass difference is smaller, as occurs in some models of non-thermal production then it might be possible to determine the wino nature of the LSP by looking for charged tracks, as recently studied in \cite{Yanagida}. More generally, several LHC signatures will depend on the mass and type of the LSP, so we expect that careful studies will be able to test whether a candidate seen in indirect data is also present in LHC data.

\section*{Acknowledgments}
We would like to thank Bobby Acharya, Joakin Edsj\"o, Igor Moskalenko, Alexey Petrov, Pierre Salati, Andrew Strong, and Gary Watson for useful conversations.  Thanks also to Doug Finkbeiner, Greg Dobler and especially Dan Hooper for discussions relevant to synchrotron radiation.  SW would also like to thank the Stony Brook physics department
and the 2008 Simons' Workshop in Mathematics and Physics for hospitality and partial financial assistance. The work of AP is supported by the Michigan Center for Theoretical Physics (MCTP) and NSF CAREER grant NSF-PHY-0743315.  The work of PG, GK, DP, and SW is supported  by the MCTP and the DOE under grant DE-FG02-95ER40899.


\end{document}